\documentclass[11pt]{article}
\usepackage{graphicx}
\usepackage{amssymb}
\usepackage{epstopdf}

\def\Eq#1{Eq.~(\ref{#1})}
\newcommand{\dslash}{{\not{{\hspace{-0.25em}}}} {\partial}}
\newcommand{\aslash}{{\not{{\hspace{-0.25em}}}} {\hspace{-0.25em}}A}

\newcommand{\jpi}{J_{{\pi}}}
\newcommand{\jsigma}{J_{{\sigma}}}
\newcommand{\W}{\mathcal{W}}

\newcommand{\trc}{\rm{Tr_c}}
\newcommand{\Tint}{\int_0^{{\infty}} {\frac{dT}{T}} exp  \left\{
-{\frac{m^2}{2}}T \right\}}
\newcommand{\pathint}{\int_{x,{\psi}} exp  \{ -S_0 \}  <{\W}[A]>_A}
\newcommand{\vsigma}{V_{{\sigma}}}
\newcommand{\vpi}{V_{{\pi}}}
\newcommand{\vzero}{V_0}
\newcommand{\vscalar}{V_s}

\begin{document}
\def\thepage{}
\begin{titlepage}

\title{Sigma Mass in the Large $N$ Limit of QCD} \author{\\
Vikram Vyas{\footnote{Associate, Abdus Salam International Centre for
Theoretical Physics, Trieste, Italy.}}\\
\textit{Physics Department} \\
\textit{St. Stephen's College, Delhi University}\\
Email: vikram@ictp.trieste.it } 
\date{}
\maketitle

\begin{abstract}
  We study the leading contribution to the scalar and the pseudoscalar
  two-point function in the large $N$ expansion of QCD using the worldline
  techniques. We find that in this limit there exists a relationship between these
  two Green's functions which implies that
  for every massive pseudoscalar meson there exists a scalar meson of identical mass.  
  This is true even when chiral symmetry is spontaneously broken. The
  relationship between the Green's function further suggests that in the planar limit 
  the sigma mass must be identical to the eta mass. We also discuss
  the relevance of these results for the quenched lattice QCD simulations and
  for hadron phenomenology.
\end{abstract}
\end{titlepage}

\renewcommand{\thepage}{\arabic{page}}
\setcounter{page}{1}

\section{Introduction} \label{sec:1}

It is an abiding hope that a $SU(N)$ gauge theory is exactly solvable in the limit in which $N$ goes to infinity, perhaps through some dual string description~\cite{tHooft, polybook, witteN}, and that this planar limit provides an appropriate starting point for an approximate solution of QCD in inverse powers of $N$. Although this hope has still not been realised, the analysis of the planar limit does show that it captures many of the gross features of meson phenomenology. An important result that further strengthens our belief in the large $N$ description of QCD is the theorem by Coleman and Witten~\cite{colemanWitten}, which states that, under certain reasonable assumptions, chiral symmetry must be spontaneously broken in the planar limit.

Thus, we expect that the pion must be massless in the planar limit of chiral QCD. Can we say some thing as well about its scalar partner, the sigma, in this limit?  The main aim of this paper is to try and answer this question by combining the worldline techniques {\cite{schubertPhyRep}} with the known analytic structure of the meson two-point functions in the planar limit.  In the planar limit, heuristically speaking, the quark degrees of freedom can be treated using relativistic quantum mechanics and therefore the motion of the quarks constituting a meson can be described by a path integral. As we will see, it is for this reason that the worldline techniques turn out to be convenient in exploring the planar limit.

The outline of the paper is as follows. In the next section we will use the known
worldline representation of the one-loop fermionic effective action
{\cite{schubert1, DHoker}} to develop  vertex functions for the scalar and the
pseudoscalar mesons.  Using
these vertex functions we obtain, in section 3, a relationship between the
scalar and the pseudoscalar two-point function. This relationship implies that for every massive pseudoscalar meson there exists, in the planer limit, a scalar meson with identical mass.
Under plausible assumptions, the same relationship implies that the sigma mass must be identical to the mass of the second lightest pseudoscalar meson, namely the eta. In the last section of the
paper we discuss the relevance of our results to the quenched simulation of
lattice QCD and to hadron phenomenology.

\section{Pion and Sigma Vertex Functions in the Planar Limit} \label{sec:2}

Consider the partition function of a $SU ( N )$ gauge theory in Euclidean space with one flavour of
quark\footnote{We consider only one flavour, as the flavour degrees of freedom do not play any dynamical role in the planar limit.} whose mass is $m$
\begin{equation}
  Z [ J ] = \int_{A, \Psi} \exp \{ - S_{} [ A ] \} \exp \left\{ \int_x
  \bar{\Psi} ( - i \dslash - \aslash - im - i \jsigma - \gamma_5 \jpi ) \Psi
  \right\} \label{zym},
\end{equation}
where $A$ is the gauge field, $S[A]$ is the Yang-Mill action for the gauge field, while $\Psi$ is the quark field and $\jsigma$ and $\jpi$ acts as the sources for the scalar and the pseudoscalar mesons. A convenient order parameter for chiral symmetry is
\begin{equation}
  < 0| \bar{\Psi} ( y ) \Psi ( y ) |0 > = \left( i \frac{\delta}{\delta
  \jsigma ( y )} \ln Z [ \jsigma, \jpi ] \right)_{J = 0} , \label{psibarpsi}
\end{equation}
and we would like to obtain a worldline representation for it. To do so we first write the partition function in terms of the fermionic determinant,
\begin{equation}
  Z [ J ] = \int_A \exp \{ - S_{} [ A ] \} \exp \{ - \Gamma [ J, A ] \}
\end{equation}
and note that the fermionic effective action
\begin{equation}
  - \Gamma [ J, A ] = \ln\det ( - i \dslash - \aslash - im - i \jsigma -
  \gamma_5 \jpi ) 
\end{equation}
has a worldline representation\cite{schubert1, DHoker} given by
\begin{equation}
  \label{WLeffA} 
  \Gamma [ J, A ] = \int_0^{\infty} \frac{dT}{T} \exp \{ - m^2
  \frac{T}{2} \} \int_{x, \psi_a} \exp \{ - ( S_0 + S [ J ] ) \} \W [ A ].
\end{equation}
 The worldline actions $S_0$ and $S[J]$ are given by
\begin{eqnarray}
  S_0 & = & \int_0^T d \tau \{ \frac{\dot{x}^2}{2} + \frac{1}{2} \psi_{\mu}
  \dot{\psi_{\mu}} + \frac{1}{2} \psi_5 \dot{\psi_5} + \frac{1}{2} \psi_6
  \dot{\psi_6} \},  \label{s0}\\
  S [ J ] & = & \int_0^T d \tau \{ \frac{1}{2} J_{\pi}^2 + i \psi_{\mu} \psi_5
  \partial_{\mu} J_{\pi} + mJ_{\sigma} + \frac{1}{2} J_{\sigma}^2 + i
  \psi_{\mu} \psi_6 \partial_{\mu} J_{\sigma} \},  \label{sj}
\end{eqnarray}
while the Wilson loop for a spin-half particle is defined as
\begin{equation}
\W [ x, \psi_{\mu} ; A ] = \trc \hat{P}\exp \left\{ i \int_0^T d \tau \{
		\dot{x}_{\mu} A_{\mu} - \frac{1}{2} \psi_{\mu} F_{\mu \nu} \psi_{\nu} 	\right\} . \label{fwl}
\end{equation}
In the path integral (\ref{WLeffA}), a path of a quark is specified by
four bosonic coordinates $x_\mu( \tau )$ and by six fermionic, or
the anti-commuting, coordinates $\psi_a( \tau )$ (please see Ref.~\cite{DHoker} for details.)
In terms of $\Gamma[J, A]$ the order parameter $<\bar{\Psi}\Psi>$ can be written as
\begin{eqnarray}
< 0| \bar{\Psi}(y) \Psi ( y ) |0 > & = & \frac{1}{Z [ 0 ]} \int_A \exp \{ - S_{}
 						 [ A ] \}  \nonumber \\
					&  & \times  \exp \{ - \Gamma [ A ] \} 
					\left( - i \frac{\delta}{\delta \jsigma (y)} \Gamma [ J, A ] \right)_{J = 0}.
\end{eqnarray}
If we restrict ourselves to the leading term in the large $N$ expansion then we can write this as
\begin{eqnarray}
< 0| \bar{\Psi}(y) \Psi ( y ) |0 > & = & \frac{1}{Z_{YM}} \int_A \exp \{ - S_{} [ A ] \}  
					\left( - i \frac{\delta}{\delta \jsigma (y)} \Gamma [ J, A ] \right)_{J = 0} \label{quenched} \\
					Z_{YM} &=& \int_{A} \exp \{ - S_{} [ A ] \} .
\end{eqnarray}
It is worth noting that this would also be the expression for the order parameter in the quenched approximation of lattice QCD. By substituting~(\ref{WLeffA}) in the above expression we readily obtain the worldline representation for the order parameter
\begin{equation}
  < 0| \bar{\Psi}(y) \Psi ( y ) |0 > = \int_0^{\infty} \frac{dT}{T} \exp \{ m^2
  \frac{T}{2} \} \int_{x, \psi_a} \exp \{ - S_0 \} < \W [ A ] >_A \vsigma ( y
  ),
\end{equation}
where the expectation value $<\W>_A$ is given by
\begin{equation}
\label{defWavg}
<\W[A]>_A = \frac{1}{Z_{YM}}\int_A \exp \{-S_{}[A]\} \W[A].
\end{equation}
and we have defined the scalar vertex function $V_{\sigma} ( y )$ as
\begin{equation}
  \vsigma ( y ) = \left( i \frac{\delta}{\delta \jsigma ( y )} S [ J ]
  \right)_{J = 0}.
\end{equation}
The explicit form of the vertex function can be obtained by using (\ref{sj}) and is
\begin{equation}
  \label{Vsigma}
   \vsigma ( y ) = i \int_0^T d \tau \left\{ m \delta ( x ( \tau
  ) - y ) + i \psi_{\mu} ( \tau ) \psi_6 ( \tau ) \partial_{\mu} \delta ( x (
  \tau ) - y ) \right\} .
\end{equation}
It will be convenient to write the above vertex function
as
\begin{equation}
  \vsigma ( y ) = m \vzero ( y ) + V_s ( y ),
\end{equation}
where
\begin{eqnarray}
  \vzero ( y ) & = & i \int_0^T d \tau \delta ( x ( \tau ) - y ), \\
  V_s ( y ) & = & - \int_0^T d \tau \psi_{\mu} ( \tau ) \psi_6 ( \tau )
  \partial_{\mu} \delta ( x ( \tau ) - y ) . 
\end{eqnarray}
Following a similar set of arguments one can write a worldline
representation for the scalar two-point function, defined as
\begin{equation}
  \label{sigmaProp} 
  \Delta_{s} ( y_1 - y_2 ) = < 0| \bar{\Psi}(y_2) \Psi ( y_2 ) \bar{\Psi}(y_1) \Psi ( y_1 ) |0 >,
\end{equation}
in terms of the scalar vertex function\footnote{In writing the two-point function in terms of $\vsigma ( y)$  we are neglecting a contact term as it will not play any role in our discussion.} 
\begin{equation}
  \label{wlsigmaprop}
   \Delta_{s} ( y_1 - y_2 ) = < \vsigma ( y_1 )
  \vsigma ( y_2 ) >_{wl},
\end{equation}
where the worldline average of any worldline functional, say $F$, is given by
\begin{equation}
  \label{defWLavg}
   < F >_{wl} = \Tint \pathint F.
\end{equation}
Let us next consider the pseudoscalar two-point function,
\begin{eqnarray}
  \Delta_{ps} ( y_1 - y_2 ) & = & < 0| \bar{\Psi}(y_1) \gamma_5 \Psi ( y_1 )
  \bar{\Psi}(y_2) \gamma_5 \Psi ( y_2 ) |0 > . \label{pionProp}
\end{eqnarray}
It too has a worldline representation in terms of the pseudoscalar vertex function, defined as
\begin{eqnarray}
  V_{\pi} & = & \left( i \frac{\delta}{\delta \jpi ( y )} S [ J ] \right)_{J =
  0}, \\
  V_{\pi} & = & - \int_0^T d \tau \psi_{\mu} \psi_5 \partial_{\mu} \delta ( x
  ( \tau ) - y ) ,
\end{eqnarray}
and is given by
\begin{equation}
  \Delta_{ps} ( y_1 - y_2 ) = < V_{\pi} ( y_1 ) V_{\pi} ( y_2 ) >_{wl} .
\end{equation}
For studying the chiral limit it will be convenient to write
the above vertex functions in momentum space. The sigma vertex function in
the momentum space is
\begin{equation}
  \label{scalarVertexInK} 
  \vsigma ( k ) = m \vzero ( k ) + V_s ( k )
\end{equation}
where,
\begin{eqnarray}
  \vzero ( k ) & = & i \int_0^T d \tau \exp \{ ik.x ( \tau ) \}, \\
  V_s ( k ) & = & - i \int_0^T d \tau k. \psi ( \tau ) \psi_6 ( \tau ) \exp \{
  ik.x ( \tau ) \} . 
\end{eqnarray}
Similarly the momentum space pion vertex function
is
\begin{equation}
  \label{vpik} 
  \vpi ( k ) = - i \int_0^T d \tau k. \psi ( \tau ) \psi_5 ( \tau )
  \exp \{ ik.x ( \tau ) \} .
\end{equation}
Coming back to the worldline representation of the order
parameter for chiral symmetry, using the vertex functions described
above,
\begin{eqnarray}
  < 0| \bar{\Psi} \Psi ( y ) |0 > & = & m < \vzero ( y ) >_{wl} + < V_s
  >_{wl}, \\
  & = & m < i \int_0^T d \tau \delta ( x ( \tau ) - y ) >_{wl} \nonumber\\
  &  & + < - \int_0^T d \tau \psi_{\mu} ( \tau ) \psi_6 ( \tau )
  \partial_{\mu} \delta ( x ( \tau ) - y ) >_{wl} . 
\end{eqnarray}
The second term in the above equation vanishes as the
integrand is odd in the Grassmann variable $\psi_6$ and the order parameter
can be written as
\begin{equation}
  \label{wlorder} 
  < 0| \bar{\Psi} \Psi ( y ) |0 > = m < i \int_0^T d \tau \delta ( x
  ( \tau ) - y ) >_{wl} .
\end{equation}
Separating the zero mode of a path, defined
as,
\begin{eqnarray}
  x ( \tau ) & = & x_0 + \bar{x} ( \tau ) \\
  x_0 & = & \frac{1}{T} \int_0^T d \tau x ( \tau ), 
\end{eqnarray}
finally gives the worldline representation for the order
parameter as
\begin{equation}
  \label{} < 0| \bar{\Psi} \Psi ( y ) |0 > = im \int_0^{\infty} dT \exp \{ -
  \frac{1}{2} m^2 T \} \int_{\bar{x}, \psi} \exp \{ - S_0 \} < \W >_A
\end{equation}
which is identical to that obtained by Banks and Casher\cite{banks}, except the trace over gamma matrices has been replaced by the
path integral over the Grassmann variables $\psi ( \tau )$. As pointed out in
\cite{banks} the spontaneous breaking of chiral symmetry requires that in
the limit the current quark mass $m$ goes to zero
\begin{equation}
  \label{banksCasherRelation} 
  \int_{\bar{x}, \psi} \exp \{ - S_0 \} < \W >_A \sim \frac{1}{T^{1 /
  2}} .
\end{equation}

\section{Sigma Mass in the Large $N$ limit} \label{sec:3}

In the planar limit of the large $N$ expansion the analytic
structure of a two-point function is greatly simplified~\cite{witteN}. Thus the scalar two-point function~(\ref{sigmaProp}) is an infinite sum of terms made up of simple poles
\begin{eqnarray}
   \Delta_{s} ( k ) & = & \sum_{n = 1}^{\infty} \frac{G_n}{k^2 + M_{s, n}^2}. \label{largeNsTwoPoint}
\end{eqnarray}
Similarly the pseudoscalar two-point function~(\ref{pionProp}) in the planar limit is 
\begin{eqnarray}
   \Delta_{ps} ( k ) & = & \frac{F_{\pi}}{k^2 + m_{\pi}^2} + \sum_{n = 1}^{\infty}
  \frac{F_n}{k^2 + M_{ps, n}^2}, \label{largeNpsTwoPoint}
\end{eqnarray}
where we have separated the pion pole from the infinite sum,  anticipating its role as the Nambu-Goldstone boson in chiral limit.

Let us now consider the worldline representation of these two Green's functions. For the scalar two-point function we have, using~(\ref{wlsigmaprop}) and~(\ref{scalarVertexInK})
\begin{equation}
  \label{scalarWLinK} 
  \Delta_{s} ( k ) = m^2 < \vzero ( k ) \vzero ( - k ) >_{wl} +
  < \vscalar ( k ) \vscalar ( - k ) >_{wl}.
\end{equation}
In writing the above expression we have neglected the cross
term, $< \vzero \vscalar >$, as it vanishes being odd in the Grassmann
variable $\psi_6$. Similarly, the worldline representation of the pseudoscalar two-point function is
\begin{equation}
  \label{twopointPS} 
  \Delta_{ps} ( k ) = < \vpi ( k ) \vpi ( - k ) >_{wl}.
\end{equation}
Next we notice that
\begin{equation}
  < \vpi ( k ) \vpi ( - k ) >_{wl} = < \vscalar ( k ) \vscalar ( - k ) >_{wl},
\end{equation}
as the right hand side can be obtained from the left hand side
by merely relabelling the Grassmann variable $\psi_6 $ as $\psi_5$ and $\psi_5$ as $\psi_6$. Using this result  in \Eq{scalarWLinK} leads to the following relationship between the scalar and the pseudoscalar two-point functions
\begin{eqnarray}
  	 \Delta_{s} ( k ) &=& m^2 < \vzero ( k ) \vzero ( - k ) >_{wl}  + \Delta_{ps}(k).
  \label{relation}
\end{eqnarray}
If we add to this our knowledge of the analytic structure of the pseudoscalar two-point function~(\ref{largeNpsTwoPoint}) then we obtain
\begin{eqnarray}
\Delta_{s} ( k ) & = &m^2 < \vzero ( k ) \vzero ( - k ) >_{wl} + \frac{F_{\pi}}{k^2 +
  m_{\pi}^2} \nonumber\\
  &  & + \sum_{n = 1}^{\infty} \frac{F_n}{k^2 + M_{ps, n}^2} .\label{poleRelation}
\end{eqnarray}
Now consider the above relationship in chiral limit,
namely in the limit $m$ going to zero. In chiral limit of the planar QCD we
expect that the lightest pseudoscalar meson, the pion, should become massless
being the Nambu-Goldstone boson of the spontaneously broken chiral symmetry
\cite{colemanWitten}, but (\ref{poleRelation}) would imply that in the chiral
limit there should also be a massless scalar meson. In what follows we will assume that there is no massless scalar in the planar limit. The main reasons for making this assumption is that, if the planar limit is a good approximation to QCD, as it seems to be for the gross features of the meson phenomenology, and since there are no light scalar mesons in the QCD spectrum we do not expect the same in the planar limit (but see~\cite{haridass} for the subtleties of the sigma mass in the spontaneously broken theories.) Then the only way of reconciling the relation~(\ref{poleRelation}) with the assumed absence of a massless scalar is to require that the worldline average $<V_0(k)V_0(-k)>_{wl}$ behaves in the following manner
\begin{equation}
  \lim_{m \rightarrow 0} m^2 < \vzero ( k ) \vzero ( - k ) > = -
  \frac{F_{\pi}}{k^2} + H ( k ),
\end{equation}
where $H ( k )$ is some function which is analytic at $k = 0$.
If we combine this with the analytic structure of the scalar
two-point function in the planer limit~(\ref{largeNsTwoPoint}) then we get the following relationship
\begin{eqnarray}
  \sum_{n = 1}^{\infty} \frac{G_n}{k^2 + M_{s, n}^2} & = & H ( k ) + \sum_{n =
  1}^{\infty} \frac{F_n}{k^2 + M_{ps, n}^2} . 
\end{eqnarray}
The unknown function $H ( k )$ must
be analytic at $k = 0$ and for non-zero values of $k$ it can at most have singularities in the form of simple poles with positive residue.  Thus in the planar limit for every massive pseudoscalar meson there exists a scalar meson of \textit{identical} mass. If it does turn out that $H ( k )$ is an
entire function, then we can make a stronger prediction that the sigma
mass, the mass of the lightest scalar particle, must be identical to the mass
of the second lightest pseudoscalar meson. In other words in the planar limit
\begin{equation}
  \label{result} M_{\sigma} = M_{\eta} .
\end{equation}
It is natural to ask to what relevance this analysis has for the real world  hadron phenomenology. We discuss this issue in the next section.

\section{Conclusions}

The planar limit of QCD is quite similar to the quenched approximation made in
the lattice QCD, in both the approximations one ignores internal fermionic
loops. Of course, the planar limit is not simply QCD with internal fermionic
loops removed, it is the limit in which only a particular subset of Feynman
diagrams contribute to the Green's functions, while in the quenched QCD the
contributions from all the diagrams, including the non-planar diagrams, are included as
long as the quark lines appear only at the boundary of the diagram. Still, in
the quenched approximation we expect that the contribution of the non-planar
gluonic diagrams should go as $\frac{1}{N^2}$ with $N = 3$. Therefore, the above
conclusions obtained in the planar limit should be valid to a good
approximation even in quenched QCD. In particular, if our assumption about $H
( k )$ being an entire function is true then {\Eq{result}} should be
approximately valid even in the quenched lattice QCD simulation.

Finally, let us discuss the relevance of our results for hadron phenomenology. The expressions for the worldline vertex functions, and the relation
(\ref{relation}), are valid only in the planar limit and we expect that the corrections to
them should be of the order of $\frac{1}{N^2}$. Also, once we go beyond the planar limit
the flavour quantum numbers will start playing a role.  Therefore, if we confine
ourselves to singlet scalar and pseudoscalar mesons then we should see some remnant of
the degeneracy in their masses that exists in the planar limit.  In Table (\ref{default})  the masses 
of the lightest known scalar and pseudoscalar singlet mesons are tabulated using the data from reference~\cite{pdb}. We do see that for every pseudoscalar meson their exists a scalar meson  of approximately the same mass \footnote{Of course, the singlet masses are not well defined and have large width specially in the case of the sigma.}, and their small mass difference is consistent with our expectation that it is a $\frac{1}{N^2}$ effect which should disappears in the planar limit.
\begin{table}[htdp]
\caption{Masses of Pseudoscalar and Scalar Light Flavour Singlet
Mesons}
\begin{center}
\begin{tabular}{|c|c|c|c|}
  \hline
  Pseudoscalar  &  Mass (MeV)  &  Scalar &  Mass (MeV)\\
  \hline
  $\eta ( 547 )$  & 547  &  $\sigma$ &  400-1200 \\
  $\eta' ( 958 )$  &  958  &  $f_0 ( 980 )$  &  980\\
  $\eta ( 1295 )$ & 1295 &  $f_0 ( 1370 )$ & 1370 \\
  $\eta ( 1440 )$  & 1440 &  $f_0 ( 1500 )$ & 1500 \\
  -  &  - & $f_0 ( 1710 )$  &  1710 \\
  \hline
\end{tabular}
\end{center}
\label{default}{}
\end{table}%

\section*{Acknowledgement}

Part of  this work was done while I was visiting the Harish-Chandra Research Institute, Allahabad (HRI). I would like to thank the physics department at HRI and Rajesh Gopakumar in particular for making this visit possible. I would also like to thank  Madan Rao and Rajesh Gopakumar for very useful discussions.


\thebibliography{}
\bibitem{tHooft}G. 't Hooft, \textit{Nucl. Phys.} \textbf{B72} 
(1974) 461.

\bibitem{polybook} A. M. Polyakov, \textit{Gauge Fields and Strings}, 
 Harwood Academic Publisher, Chur (1987).

\bibitem{witteN}E. Witten, \textit{Nucl. Phys.} \textbf{B160} (1979) 57.

\bibitem{colemanWitten}S. Coleman and E. Witten,
\textit{Phys. Rev. Letts} \textbf{45} (1980) 100.

\bibitem{schubertPhyRep}C. Schubert, \textit{Phys. Rept.} \textbf{355} (2001) 73
hep-th/0101036.

%\bibitem{wilson74} K. Wilson, \textit{Phys. Rev.}
%\textbf{D10} (1974) 2445.

\bibitem{schubert1}M.Mondragon, L. Nellen, M.G. Schmidt, C. Schubert, \textit{Phys. Lett.}\textbf{B 351} (1995) 200 hep-th/9502125

\bibitem{DHoker} E. D 'Hoker and D. Gagn\`e, \textit{Nucl. Phys.} \textbf{B467 } (1996) 272\\
hep-th/9508131. 

\bibitem{banks} T. Banks and A. Casher,
\textit{Nucl. Phys.} \textbf{B169} (1980) 103


\bibitem{haridass} Ramesh Anishetty, Rahul Basu, N.D. Hari Dass and H.S.Sharatchandra, \textit{Int.J.Mod.Phys.} \textbf{A14} (1999) 3467 hep-th/9502003

\bibitem{pdb}D.E. Groom et al. (Particle Data Group), Eur. Phys. Jour. C15, 1 (2000) (URL: http://pdg.lbl.gov)
 \end{document}